\long\def\symbolfootnote[#1]#2{\begingroup%
\def\thefootnote{\fnsymbol{footnote}}\footnote[#1]{#2}\endgroup} 
\def\araa{ARA\&A}
\def\apj{ApJ}
\def\mnras{MNRAS}

\newcommand{\tcr}{T_{\rm cr}}
\newcommand{\reff}{R_{\rm eff}}
\newcommand{\ttcr}{\mbox{$\Pi$}}
\newcommand{\nobj}{105}
\newcommand{\msun}{\rm{M}_{\odot}}
\newcommand{\myr}{{\rm Myr}}
\newcommand{\tdyn}{T_{\rm dyn}}

\documentclass[useAMS,usenatbib, times]{mn2e}
\usepackage{amssymb,latexsym,graphicx,natbib,eufrak,times}

\title[The distinction between clusters and associations]
  {The distinction between star clusters and associations}
\author[M. Gieles \& S.F. Portegies Zwart]
  {Mark~Gieles$^{1}$ and Simon F. Portegies Zwart$^2$\\
  $^1$ Institute of Astronomy, University of Cambridge, Madingley Road, Cambridge, CB3 0HA, UK \\
 $^2$ Leiden Observatory, Leiden University, PO Box 9513, 2300 RA Leiden, The Netherlands  \\
}
\date{Accepted 2010 October 8.  Received 2010 October 6; in original form 2010 September 22}

\def\LaTeX{L\kern-.36em\raise.3ex\hbox{a}\kern-.15em
    T\kern-.1667em\lower.7ex\hbox{E}\kern-.125emX}

\begin{document}         

\maketitle
\begin{abstract}
In Galactic studies a distinction is made between (open) star clusters
and associations. For barely resolved objects at a distance of several
Mpc this distinction is not trivial to make.  Here we provide an
objective definition by comparing the age of the stars to the
crossing time of nearby stellar agglomerates. We find that a
satisfactory separation can be made where this ratio equals unity.
Stellar agglomerates for which the age of the stars exceeds the
crossing time are bound, and are referred to as star clusters.
Alternatively, those for which the crossing time exceeds the stellar
age are unbound and are referred to as associations. This definition
is useful whenever reliable measurements for the mass, radius and age
are available.
\end{abstract}

Star forming galaxies consist of field stars, associations and star
clusters. The distinction between star clusters and associations is
not clearly defined. \citet{amb47} introduced the term association in
reference to loose agglomerates and he pointed out in subsequent
studies that it is unlikely that they are bound by their own
gravity \citep[see also][]{1964ARA&A...2..213B}. When objects are
classified as associations, it is generally not known whether the
origin of the classification (e.g. based on the binding energy) can be
attributed to the process of formation or the evolution.  It has been
posed, and it is often quoted, that the majority of stars form in star
clusters and that there is a high rate of early cluster
disruption \citep[e.g.][]{1998MNRAS.300..200K,
2003ARA&A..41...57L}. But if the star formation process is
hierarchical then only a small fraction of the newborn stars reside in
agglomerates that satisfy the conditions necessary to be bound by
self-gravity at formation \citep[e.g.][]{2008ApJ...672.1006E,
2010MNRAS.tmpL.143B}. When observational samples of star clusters are
used to support either one of the above scenarios it is vital to know
how star clusters are separated from associations.  Here we provide a
definition of the distinction between these two classes of stellar
agglomerates.

We use the recent literature compilation of young massive clusters and
associations of \citet*[][hereafter PZMG10]{2010ARA&A..48..431P}.
This sample consists of all stellar agglomerates found in the
literature for which a value of the half-light radius $\reff$, mass
$M$, and age were determined. Their sample contains $\nobj$
agglomerates with $M\gtrsim10^4\,\msun$ and $T\lesssim100\,\myr$ in
nearby ($\lesssim10\,$Mpc) galaxies.  PZMG10 used the ratio of the
age of the stars over the crossing-time of the stars in the cluster,
$\tcr$, to distinguish star clusters from associations, where the
boundary was set at unity\footnote{In fact PZMG10 used the ratio
age$/\tdyn=3$ as the boundary, where $\tdyn$ is the dynamical
time-scale of the cluster and $\tcr/\tdyn=2\sqrt{2}\approx 2.8$ for
clusters in virial equilibrium.}. We refer to this ratio as the
 dynamical age, or $\ttcr=$\,Age$/\tcr$.

\begin{figure}
 \includegraphics[width=8.cm]{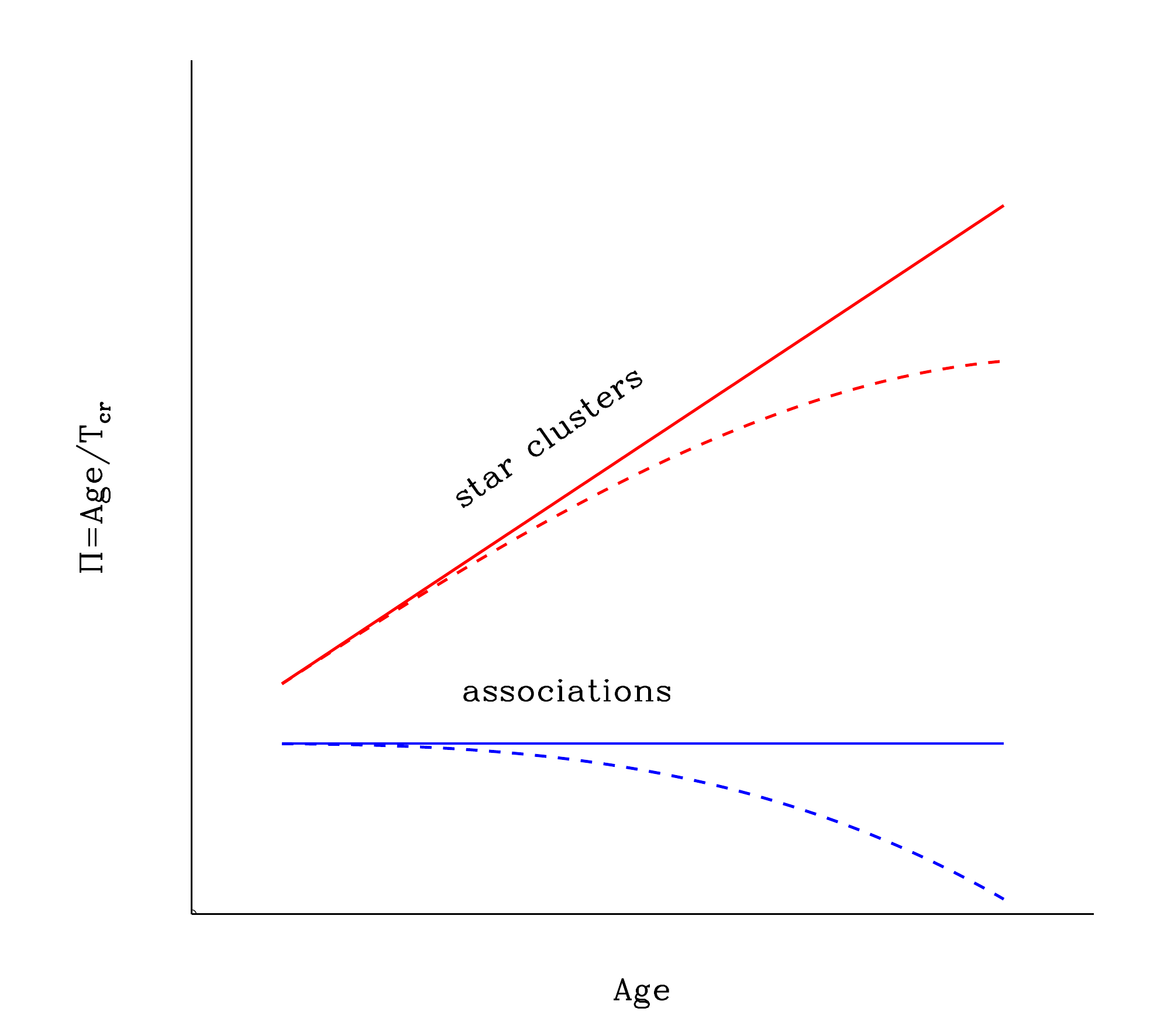} 
\caption{Schematic
    representation of the evolution of the dynamical age $\ttcr$ for
     star clusters (top lines) and associations (bottom lines). Two
     evolutionary tracks are shown for each.  The dashed line for
     clusters (top) illustrates the effect of dynamical
     expansion \citep{G10} on $\ttcr$, whereas the solid curve is
     drawn assuming that the cluster radius does not change with time.
     The full line for associations shows how $\ttcr$ evolves when
     $\tcr$ would be derived from a measured velocity dispersion
     ($\tcr\propto\reff/\sigma$, with $\sigma=\,$constant), whereas
     the dashed lines shows the evolution of $\Pi$ when
     equation~(\ref{eq:tcr}) is used to approximate $\tcr$.}
      \label{fig:ttcrschema}
\end{figure}

The boundary at $\ttcr=1$ is explicitly based on the distinction
between bound and unbound agglomerates. For expanding objects $\ttcr<1$;
the radius increases roughly proportionally with age and $\tcr$
therefore also increases. For bound objects $\ttcr>1$; we observe 
that, to first order, $\reff$ and the crossing time remain roughly
constant with time. A schematic view of the evolution of $\ttcr$ as a
function of age for star clusters and associations is shown in
Fig.~\ref{fig:ttcrschema}.

Here we define the crossing time in terms of empirical cluster
parameters that are relatively straightforward to determine
\begin{equation}
\tcr\equiv10\left(\frac{\reff^3}{GM}\right)^{1/2},
\label{eq:tcr}
\end{equation}
where $G$ is the gravitational constant.  This definition is
equivalent to equation~(11) of PZMG10 apart from a factor $2^{3/2}$ to
define the crossing time in terms of diameter instead of radius (see
footnote~1). A factor $(4/3\times16/[3\pi])^{3/2} \approx 3.4$ was used to
convert the virial radius to $\reff$. Note
that we assume a Plummer density profile and that light traces mass.

Equation~(\ref{eq:tcr}) is valid for systems in virial equilibrium. A
more general definition of $\tcr$ includes the root-mean square
velocity dispersion of the stars ($\tcr\propto\reff/\sigma$) which is
available for fewer agglomerates and at young ages the measured
$\sigma$ can be higher than the virial motion of the stars because of
orbital motions of multiples \citep*{2010MNRAS.402.1750G}. If unbound
associations expand with a constant velocity then $\tcr\propto\,$Age
and $\Pi=\,$constant (full line for associations in
Fig.~\ref{fig:ttcrschema}).  By using equation~(\ref{eq:tcr}), which
assumes virial equilibrium, we thus overestimate the increase of
$\tcr$ (underestimates $\ttcr$ at older ages) of unbound associations
thereby enlarging the difference in $\ttcr$ of bound and unbound
systems (dashed line for associations in Fig.~\ref{fig:ttcrschema}).
This definition, therefore, facilitates in making the distinction.

\begin{figure}
 \includegraphics[width=8.cm]{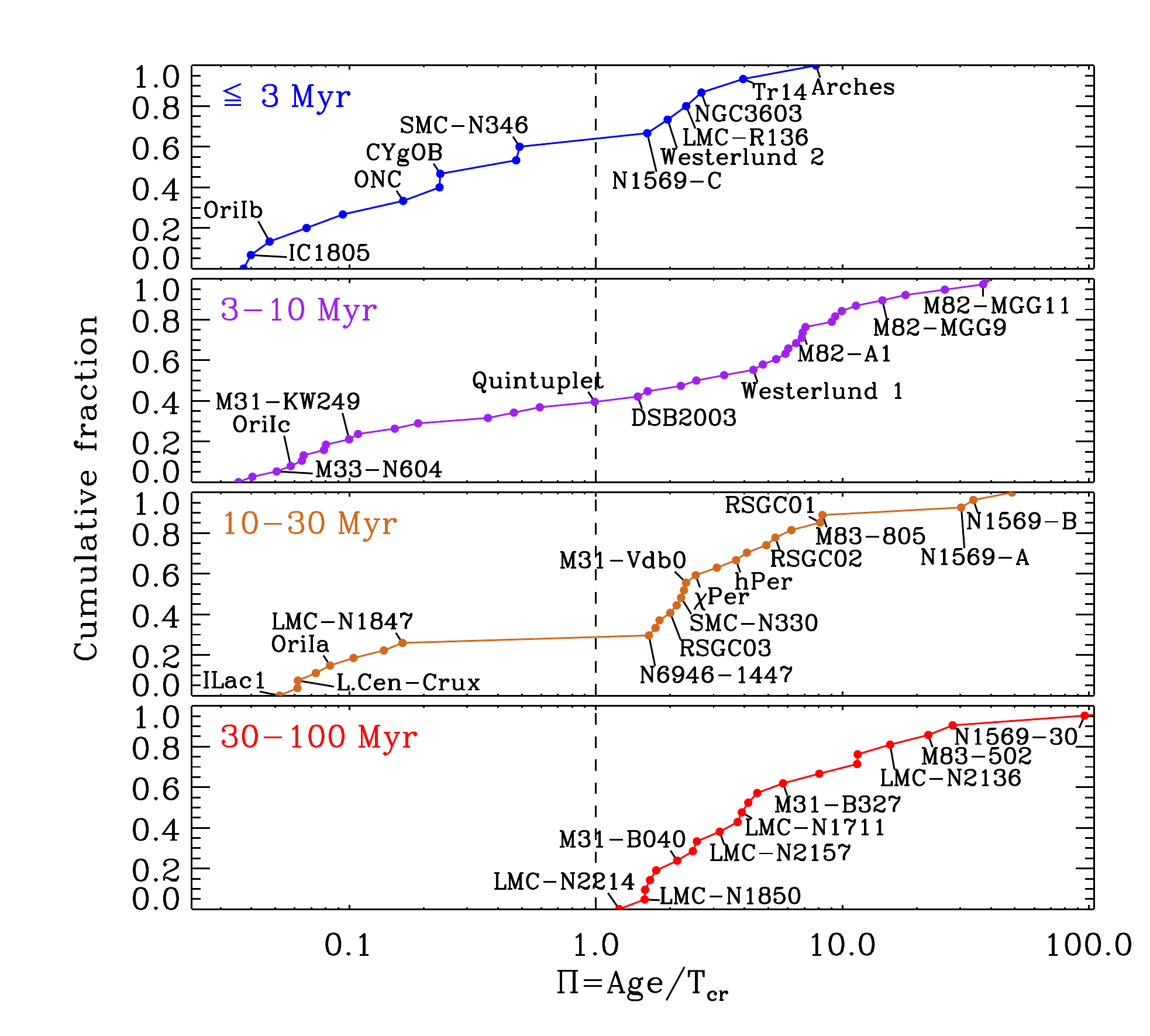}

 \caption{Cumulative distributions of \ttcr\, 
  for  nearby ($\lesssim10\,$Mpc)
     stellar agglomerates 
      (data from the compilation of PZMG10). The
     agglomerates are separated in four distinct age bin, with the
     age boundaries are indicated in the panels. Several star
     clusters and associations are identified by their common
     name. The vertical dashed line indicates the boundary value
     $\ttcr=1$.}  \label{fig:ttcrdist}
\end{figure}

In Fig.~\ref{fig:ttcrdist} we show the cumulative distribution of all
objects in different age bins.  The  top panel shows the
(cumulative) distribution of $\ttcr$ for the youngest age bin. This is a
continues distribution from $\ttcr\sim0.03$ (i.e. associations) to
$\ttcr\sim10$ (i.e. star clusters).  A similar result was recently
obtained for the surface density distribution of young stellar objects
in the solar neighbourhood \citep{2010MNRAS.tmpL.143B}.  There seems
not to be a distinct mode of star cluster formation, but rather a
smooth transition between star clusters and associations, which in
turn can be interpreted as a smooth transition between bound and
unbound objects.

The bottem panel shows that the oldest agglomerates all have
$\ttcr\gtrsim1$, which according to our definition are bound star
clusters. In this age bin there are several LMC and M31 star clusters
with $\Pi$ only slightly larger than one. This could be because most
of these clusters have rather shallow light profiles which makes
$\reff$ large compared to the core radius (PZMG10). But it can also be
that these objects are only weakly bound. The intermediate age curves
contain both associations and star clusters.  If we interpret the
curves for the different age bins as an evolutionary sequence then a
distinct gap develops between star clusters and associations around
$\sim10\,$Myr at a value of $\Pi\approx1$.  At older ages an observer
should be able to make an unambiguous distinction between an (unbound)
association and a (bound) star cluster using this straight-forward
method.  For younger ages the distributions are not separated at
$\ttcr=1$ but for the youngest (continuous) distribution it still
offers a good qualitative discrimination, as can be noted from the
labels of several well known star clusters and associations.  The
exact fraction of the newborn stars that ends up in bound star cluster
can depend on environment \citep[e.g.][]{2008ApJ...672.1006E}.
According to the definition of $\ttcr$ all agglomerates have $\ttcr=0$
when they form so it is not very meaningful to classify objects when
the star formation process is still ongoing \citep[see also
][]{2010MNRAS.tmpL.143B}.  

Our definition of the dynamical age \ttcr\, offers a dynamically
motivated and practical classification of unbound associations and
bound star clusters.  It can be applied to Galactic and
extra-galactic samples whenever estimates for masses, effective radii
and ages are available.

\vspace{-0.6cm}

\section*{Acknowledgement}
We are grateful for the KIAA in Beijing for their hospitality during
the MODEST-10 meeting where this paper was sprouted.  MG acknowledges
financial support of the Royal Society.  This work was supported by
the Netherlands Research Council NWO (via grants \#643.200.503
and \#639.073.803), by the Netherlands Research School for Astronomy
(NOVA) and by the Leids Kerkhoven-Bosscha Fonds.
\vspace{-0.6cm}

\bibliographystyle{mn2e}

\end{document}